\documentclass[aps,prl,twocolumn,preprintnumbers,superscriptaddress]{revtex4}

\usepackage{amsmath}
\usepackage{graphicx}
\usepackage{caption}
\usepackage{subfig}

\newcommand{\be}{\begin{equation}}
\newcommand{\ee}{\end{equation}}
\newcommand{\bea}{\begin{eqnarray}}
\newcommand{\eea}{\end{eqnarray}}

\def\le{\left}
\def\ri{\right}
\def\nr{N_{\rm r}}
\def\cen{\textbf{c}_*}
\def\cell{\textbf{c}}
\def\block{l}
\def\qth{q_{\rm th}}

\begin{document}
\title{Point-to-set correlations and instantons}

\preprint{MIT-CTP-4518}

\author{Sho Yaida}
\affiliation{Center for Theoretical Physics, Massachusetts Institute of Technology,
Cambridge, MA 02139, USA
}

\begin{abstract}
For a generic many-body system, we define a soft point-to-set correlation function.
We then show that this function accepts a representation in terms of an effective overlap field theory.
In particular, instantons in this effective field theory encode point-to-set correlations for supercooled liquids.
\end{abstract}
\maketitle

\newpage
Supercooled liquids exhibit dramatic slowdown with modest lowering of their temperatures, resulting in the omnipresent glass transition.
For decades, many have searched for hidden growing length scales underlying this enigmatic slowdown~\cite{BBreview}.
A promising candidate has been proposed recently by Biroli and Bouchaud~\cite{BB}: through an ingenious thought experiment, they incarnated the heuristic notion of entropic droplets~\cite{KTW} in point-to-set correlations~\cite{MSbound}.
The aim of the present paper is to further identify a class of effective field theories encoding these point-to-set correlations.

We reach our goal in three steps.
First, we define a soft point-to-set correlation function, with both an observable and a constraint expressed in terms of local overlaps [c.f.~Eq.(\ref{sPTS})].
We then apply the replica trick to eliminate a troublesome denominator.
Finally, the standard coarse-graining procedure leads us to an effective theory of an overlap field.
The resulting effective Hamiltonian [c.f.~Eq.(\ref{EFT})] encompasses the class of replica field theories, hitherto
believed to dictate the physics of supercooled liquids~\cite{DSW,CBTT}.
Furthermore, the derivation yields a microscopic justification for equating the size of instatons in this effective field theory with the point-to-set correlation length of a supercooled liquid~\cite{Franzology}.

\begin{figure*}[!ht]
  \subfloat[Pick an equilibrium configuration ${\bf X}_1$.]{\label{BB1}\includegraphics[width=.245\textwidth]{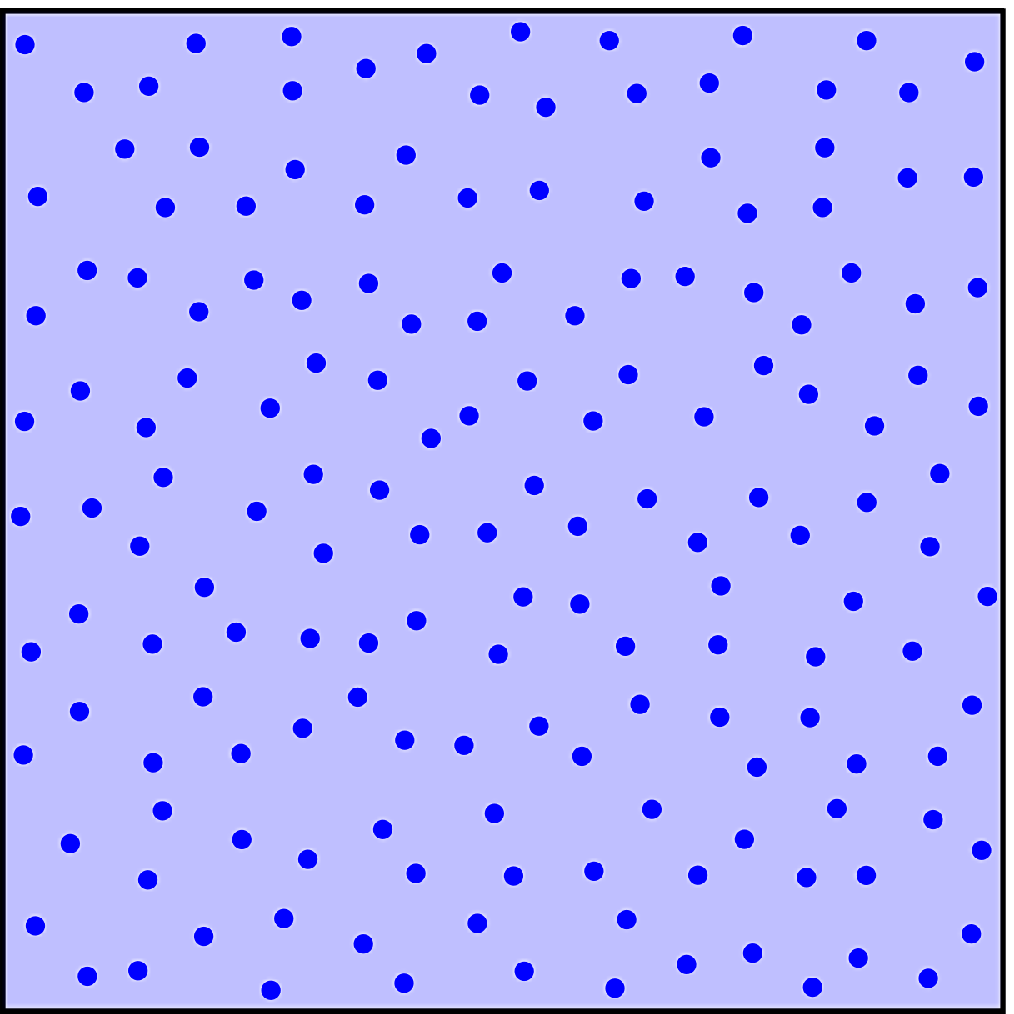}}~
  \subfloat[Fix the configuration ${\bf X}^{\textrm{out}}_1$ outside a region ${\cal R}$.]{\label{BB2}\includegraphics[width=.245\textwidth]{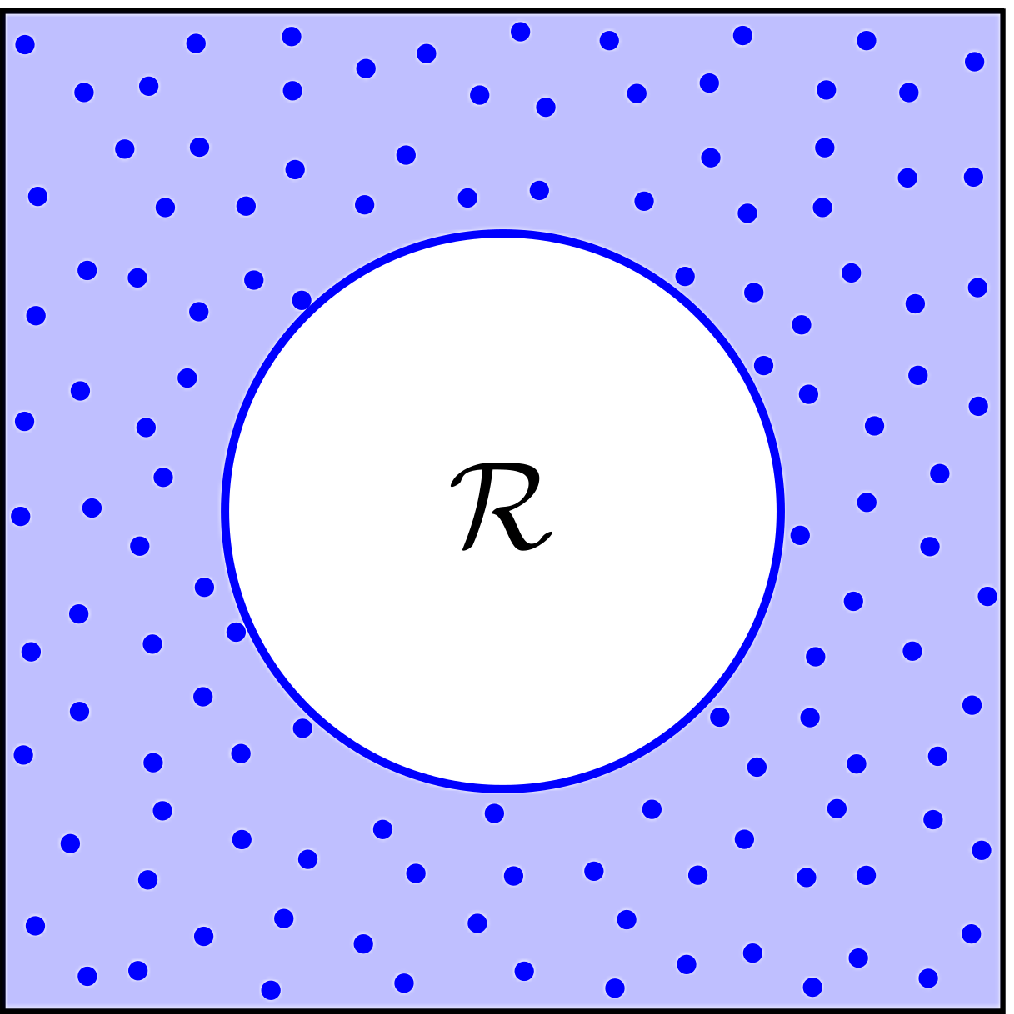}}~
  \subfloat[Pick a new conditional equilibrium configuration ${\bf X}_2$ with ${\bf X}^{\textrm{out}}_2={\bf X}^{\textrm{out}}_1$.]{\label{BB3}\includegraphics[width=.245\textwidth]{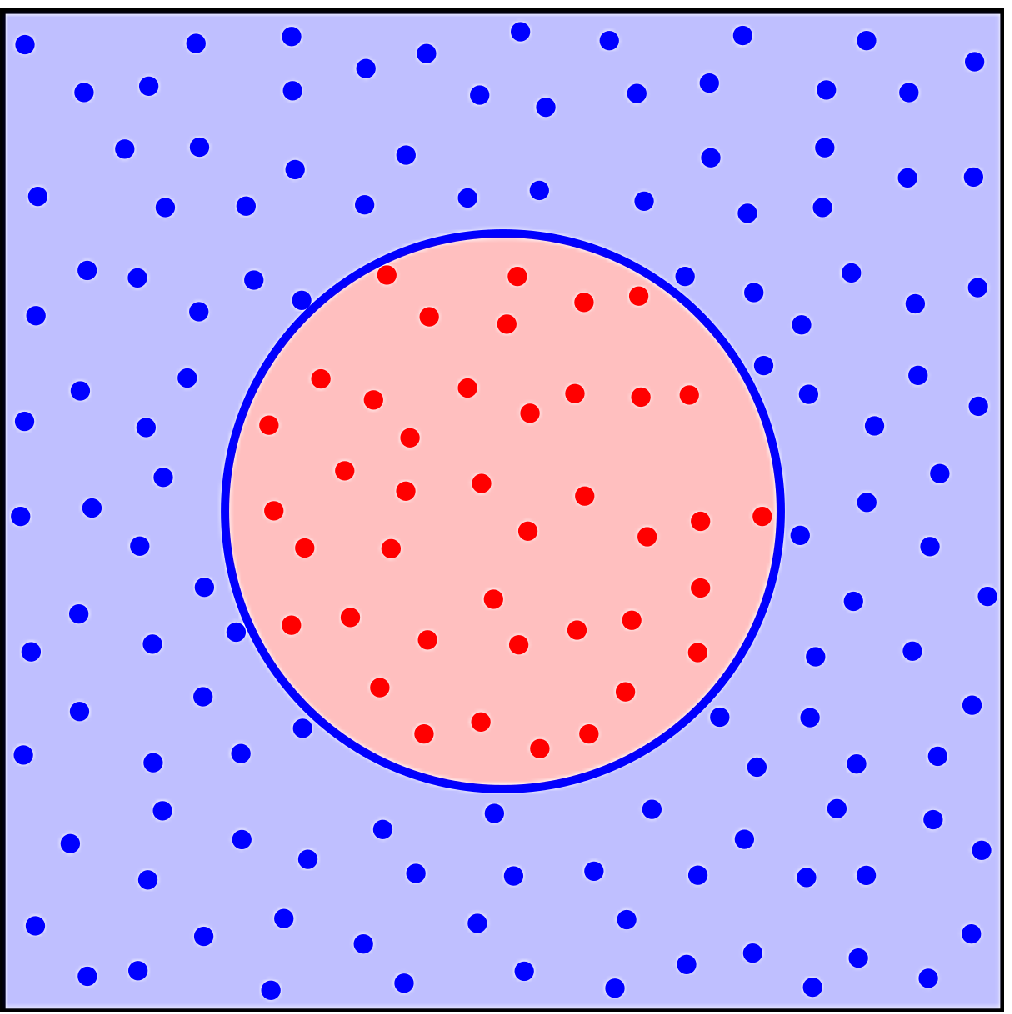}}~
  \subfloat[Compare two configurations ${{\bf X}}^{\textrm{in}}_1$ and ${{\bf X}}^{\textrm{in}}_2$ in a cell $\cen$ centered inside the region ${\cal R}$.]{\label{BB4}\includegraphics[width=.245\textwidth]{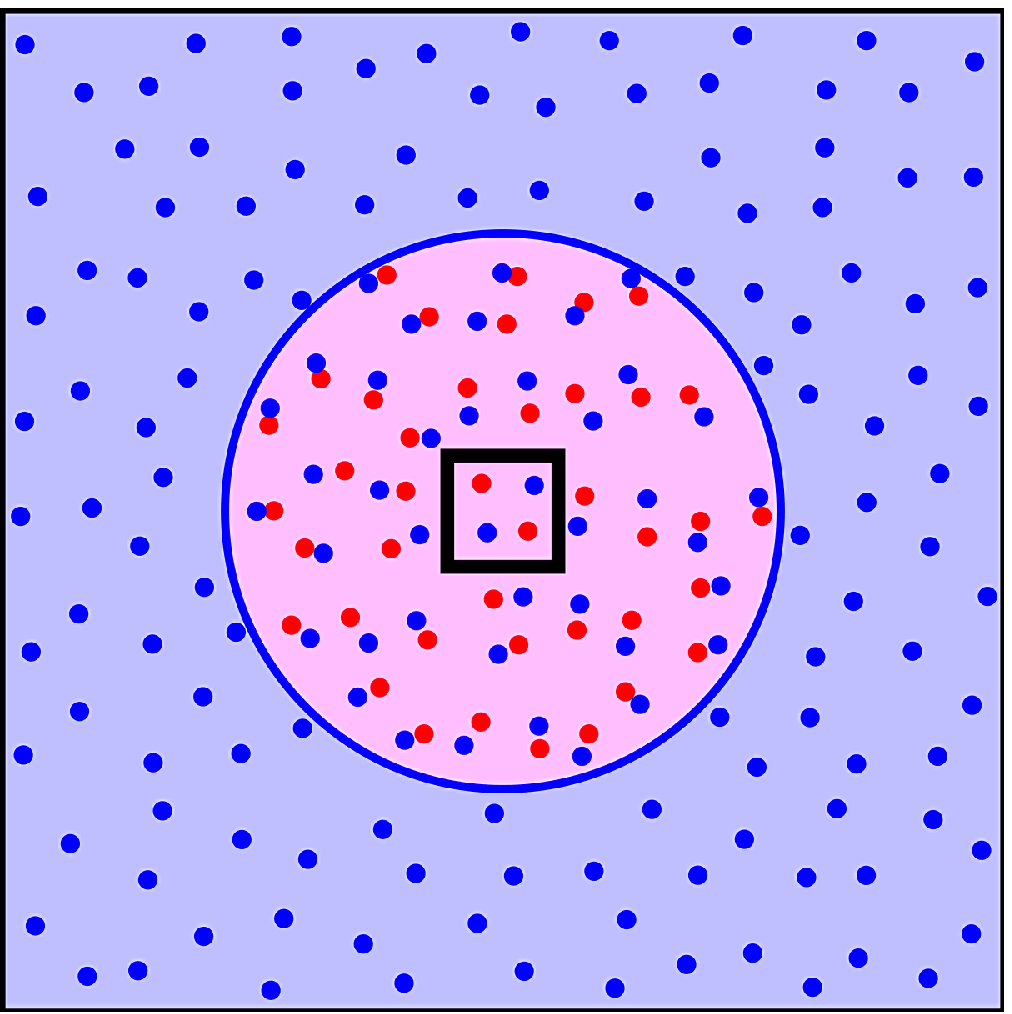}}
  \caption{The point-to-set thought experiment.}
  \label{BB}
\end{figure*}

\begin{figure*}[!ht]
  \subfloat[]{\includegraphics[width=.245\textwidth]{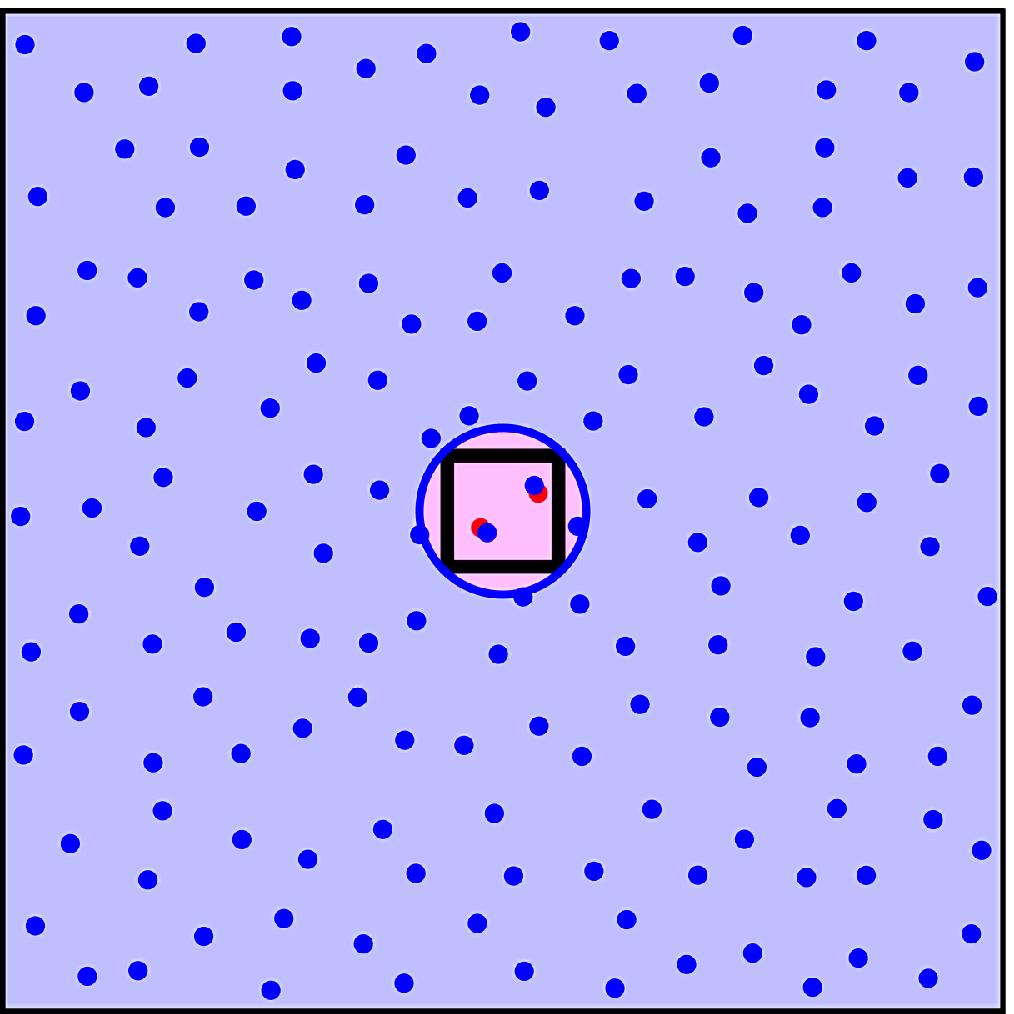}}~
  \subfloat[]{\includegraphics[width=.245\textwidth]{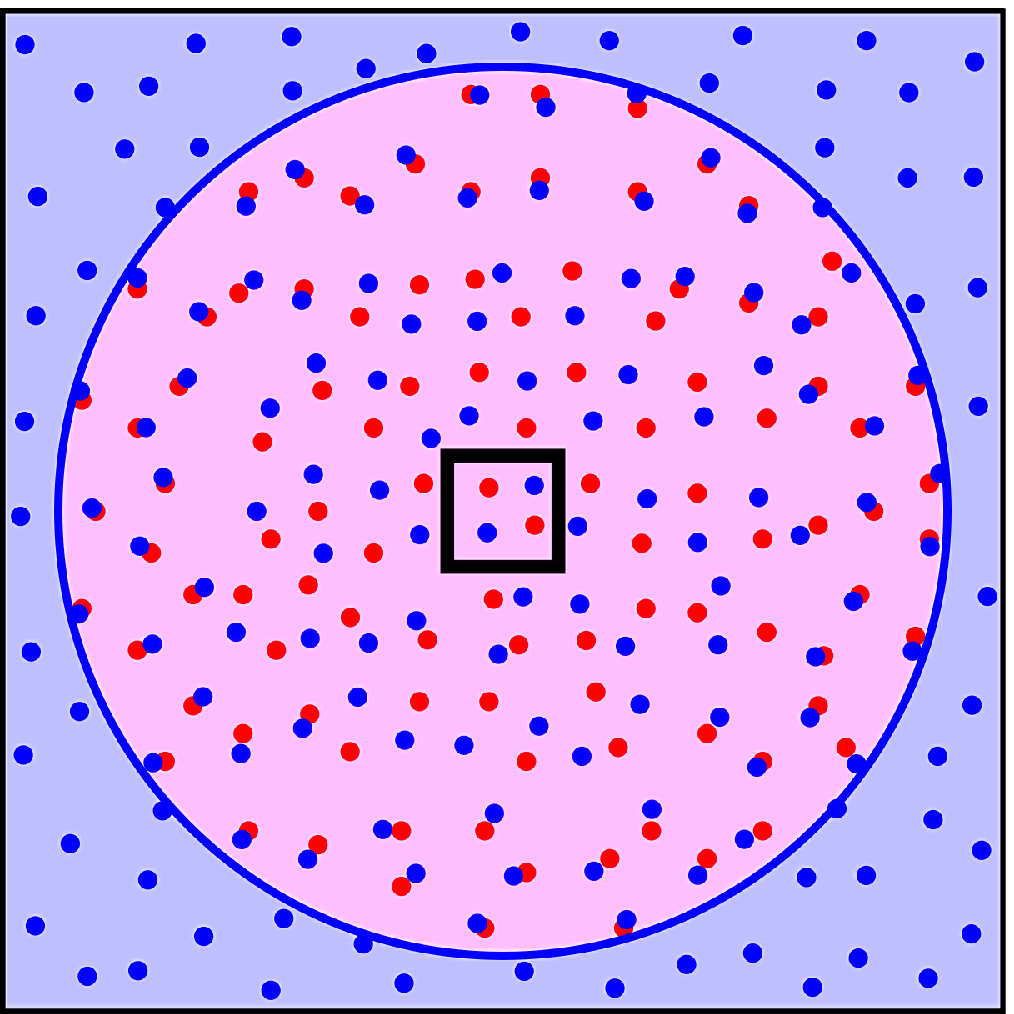}}
  \caption{(a) For sufficiently supercooled liquids, in the extreme where we take a very small region ${\cal R}$, the external force from the exterior pins the centers of vibrations of particles in the interior: we expect the original and new configurations to overlap highly. (b) In the other extreme where we take a very big region ${\cal R}$, influence from the exterior cannot survive deep inside: there, we expect two configurations to be decorrelated, resulting in low overlap.}
  \label{Extremes}
\end{figure*}

To start, let us give a mathematical expression for a conventional point-to-set correlation function, suitable for a generic many-body system consisting of $N$ particles contained in a $d$-dimensional box of volume $V$ and governed by a Hamiltonian $H$; we label each configuration by positions of the particles ${\bf X}=\le\{{\bf x}_i\ri\}_{i=1,...,N}$~\cite{details}.
Here, we shall follow the thought experiment devised in~\cite{BB}, mathematically consolidating each step.
First, we draw an equilibrium configuration ${\bf X}_1$ from a Boltzmann distribution
\be
{\cal M}({\bf X}_1)\equiv e^{-\beta H({\bf X}_1)}/Z\ ,
\ee
where the partition function $Z\equiv\int d{\bf X}e^{-\beta H({\bf X})}$ (Fig.\ref{BB1}).
Next, we fix all the particles outside some region ${\cal R}$ (Fig.\ref{BB2}), and pick a new subequilibrium configuration in the interior under the influence of the external force exerted by the fixed particles outside (Fig.\ref{BB3}).
This is realized by drawing a new configuration ${\bf X}_2$ from a conditional Boltzmann distribution
\be
{\cal M}_{\overline{\cal R}}\le({ {\bf X}}_2\Big|{\bf X}_1\ri)\equiv \frac{e^{-\beta H\le({ {\bf X}}_2\ri)}{\cal C}_{\overline{\cal R}}\le({\bf X}_2\Big|{\bf X}_1\ri)}{Z_{\overline{\cal R}}\le({\bf X}_1\ri)}
\ee
with $Z_{\overline{\cal R}}\le({\bf X}_1\ri)\equiv\int d{\bf X} e^{-\beta H({\bf X})}{\cal C}_{\overline{\cal R}}\le({\bf X}\Big|{\bf X}_1\ri)$, where the conditional function
\be\label{hard}
{\cal C}_{\overline{\cal R}}\le({\bf X}_2\Big|{\bf X}_1\ri)\equiv\delta\le({\bf X}^{\textrm{out}}_1-{\bf X}^{\textrm{out}}_2\ri)
\ee
enforces the two configurations to agree outside the region ${\cal R}$.
Then, we ask how similar this new configuration looks to the original configuration, in a cell $\cen$ centered inside the region ${\cal R}$ (Fig.\ref{BB4}).
This similarity is quantified by a local overlap $\hat{q}\le(\cen; {\bf X}_1, {\bf X}_2\ri)$, defined in the Appendix [c.f.~Eq.(\ref{overlap})].
Putting all the expressions together, our average expectation is expressed by a point-to-set correlation function
\bea\label{PTS}
&&G(\cen; \overline{\cal R})\\
&\equiv&\int d{\bf X}_1 {\cal M}({\bf X}_1)\int d{\bf X}_2 {\cal M}_{\overline{\cal R}}\le({ {\bf X}}_2\Big|{\bf X}_1\ri)\hat{q}\le(\cen; {\bf X}_1, {\bf X}_2\ri).\nonumber
\eea
For sufficiently supercooled liquids, this expectation value crossovers from a high-overlap value to a low-overlap value as we increase the size of the region ${\cal R}$ (Fig.\ref{Extremes})~\cite{BBCGV,HMR}~\cite{asymptotic}.
The crossover scale defines a point-to-set correlation length.
\captionsetup[subfigure]{textfont=normalfont,singlelinecheck=off,justification=raggedright}
\captionsetup[figure]{textfont=normalfont,singlelinecheck=off,justification=raggedright}
\captionsetup[subfigure]{textfont=normalfont,singlelinecheck=off,justification=centering}
\captionsetup[figure]{textfont=normalfont,singlelinecheck=off,justification=centering}
\captionsetup[subfigure]{textfont=normalfont,singlelinecheck=off,justification=raggedright}
\begin{figure*}[!ht]
  \subfloat[Pick an equilibrium configuration ${\bf X}_1$.]{\label{BBY1}\includegraphics[width=.245\textwidth]{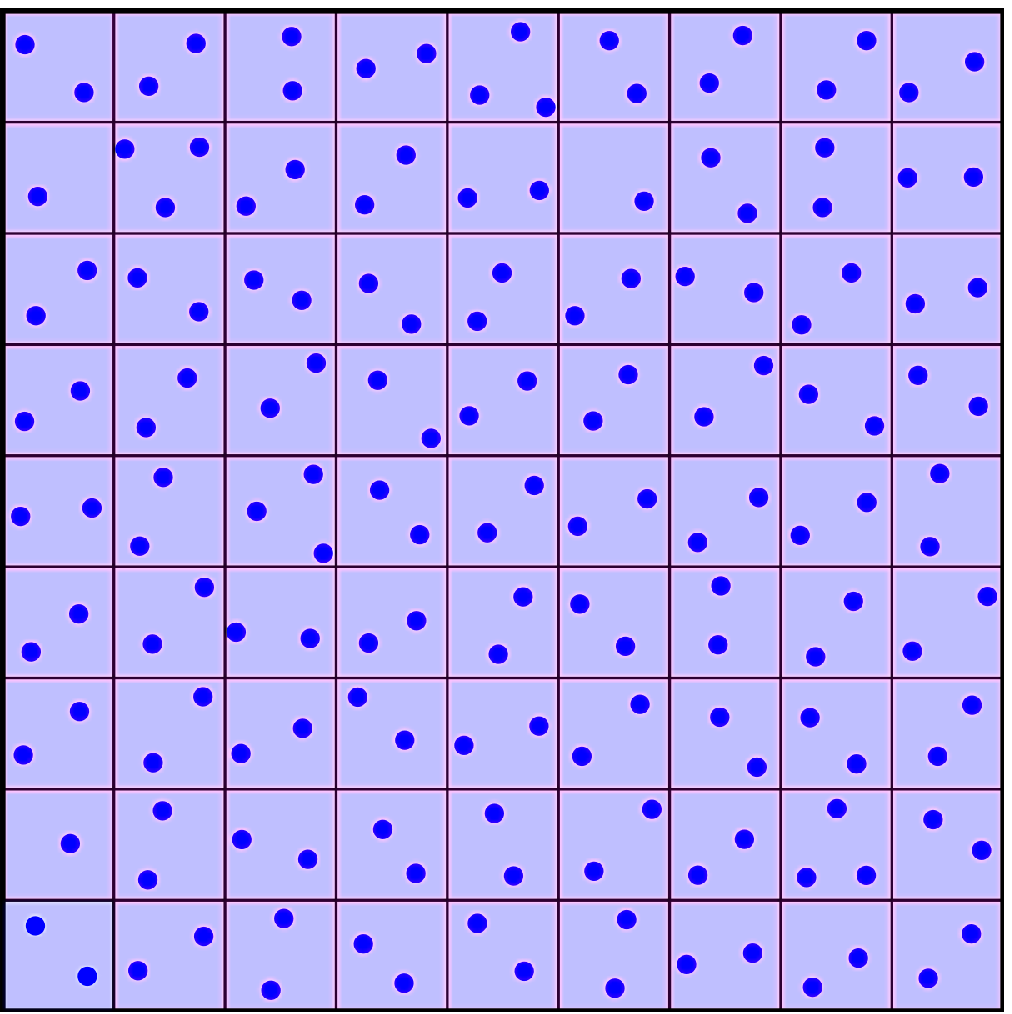}}~
  \subfloat[Fix the configuration ${\bf X}^{\textrm{out}}_1$ outside a region ${\cal R}$.]{\label{BBY2}\includegraphics[width=.245\textwidth]{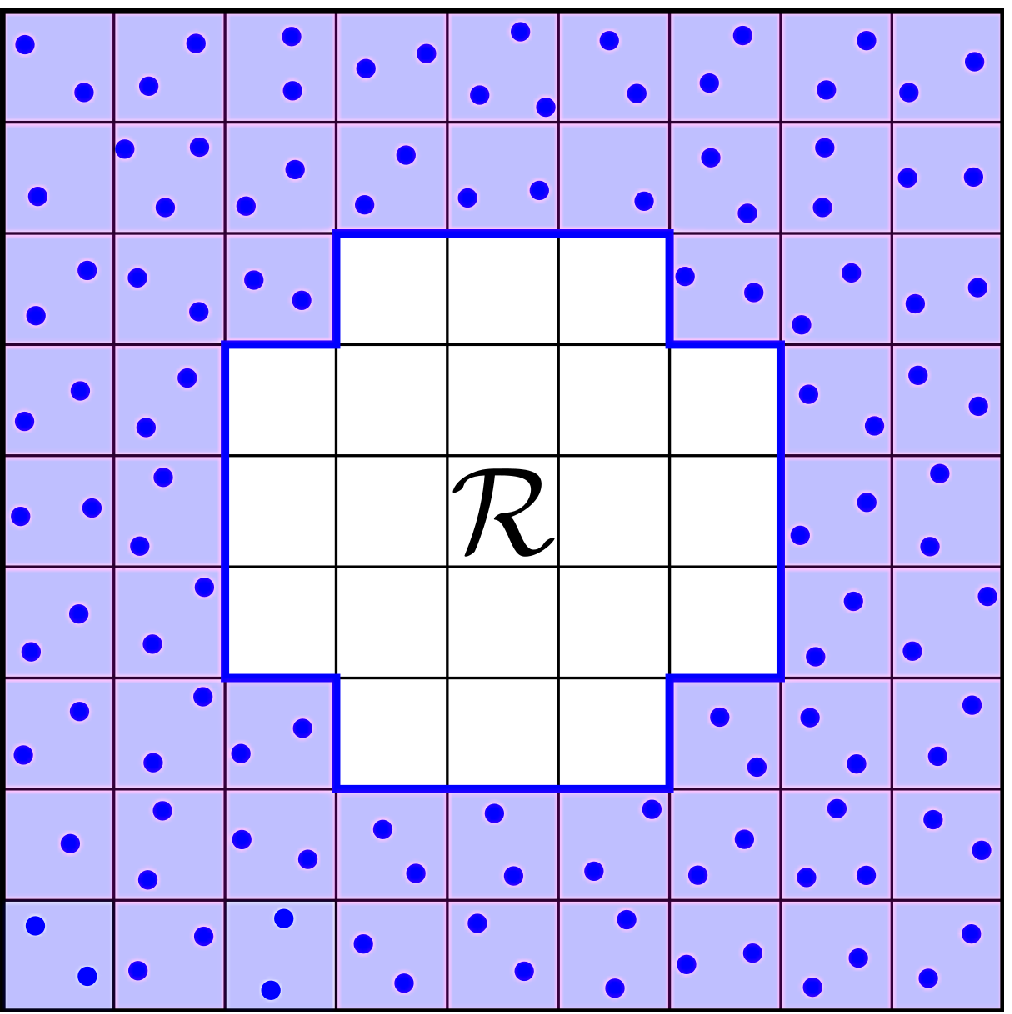}}~
  \subfloat[Pick a new softly-constrained equilibrium configuration ${\bf X}_2$ with ${\bf X}^{\textrm{out}}_2\approx{\bf X}^{\textrm{out}}_1$.]{\label{BBY3}\includegraphics[width=.245\textwidth]{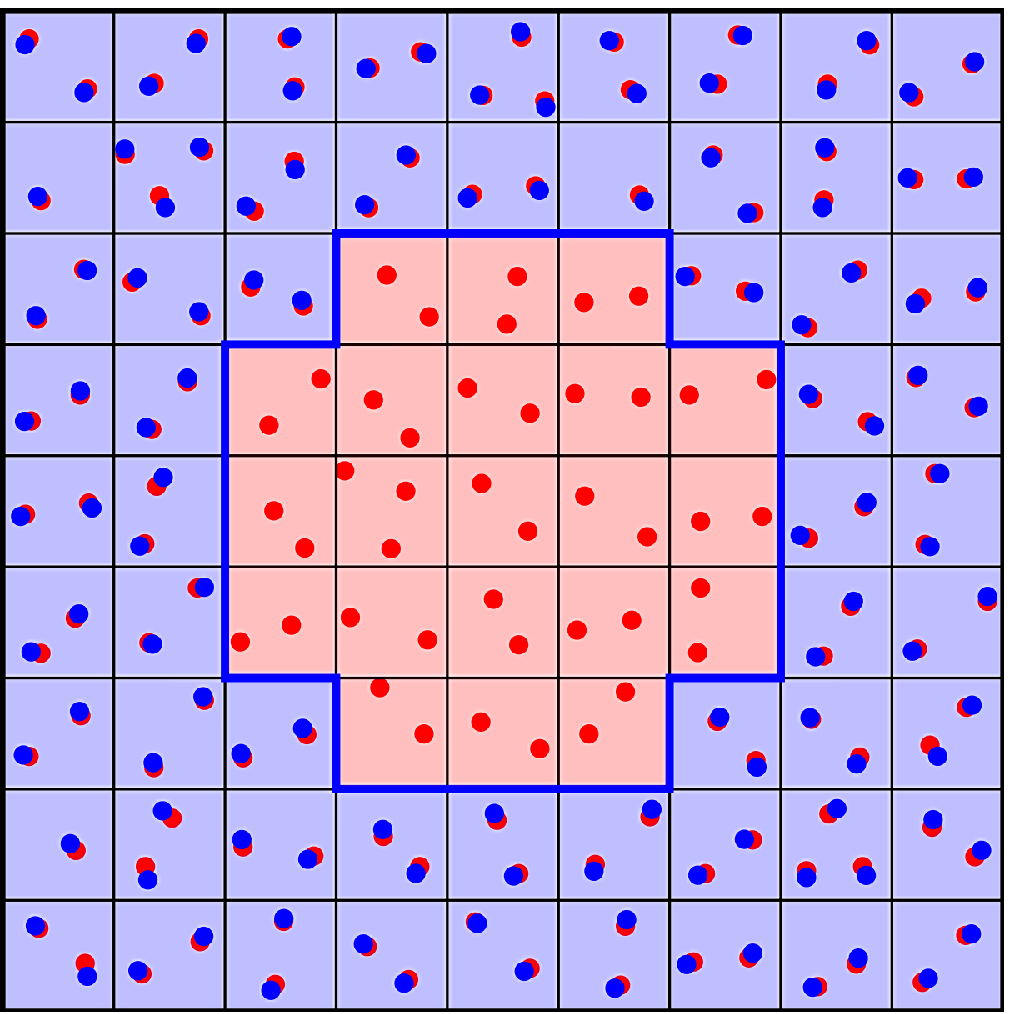}}~
  \subfloat[Compare two configurations ${{\bf X}}^{\textrm{in}}_1$ and ${{\bf X}}^{\textrm{in}}_2$ in a cell $\cen$ centered inside the region ${\cal R}$.]{\label{BBY4}\includegraphics[width=.245\textwidth]{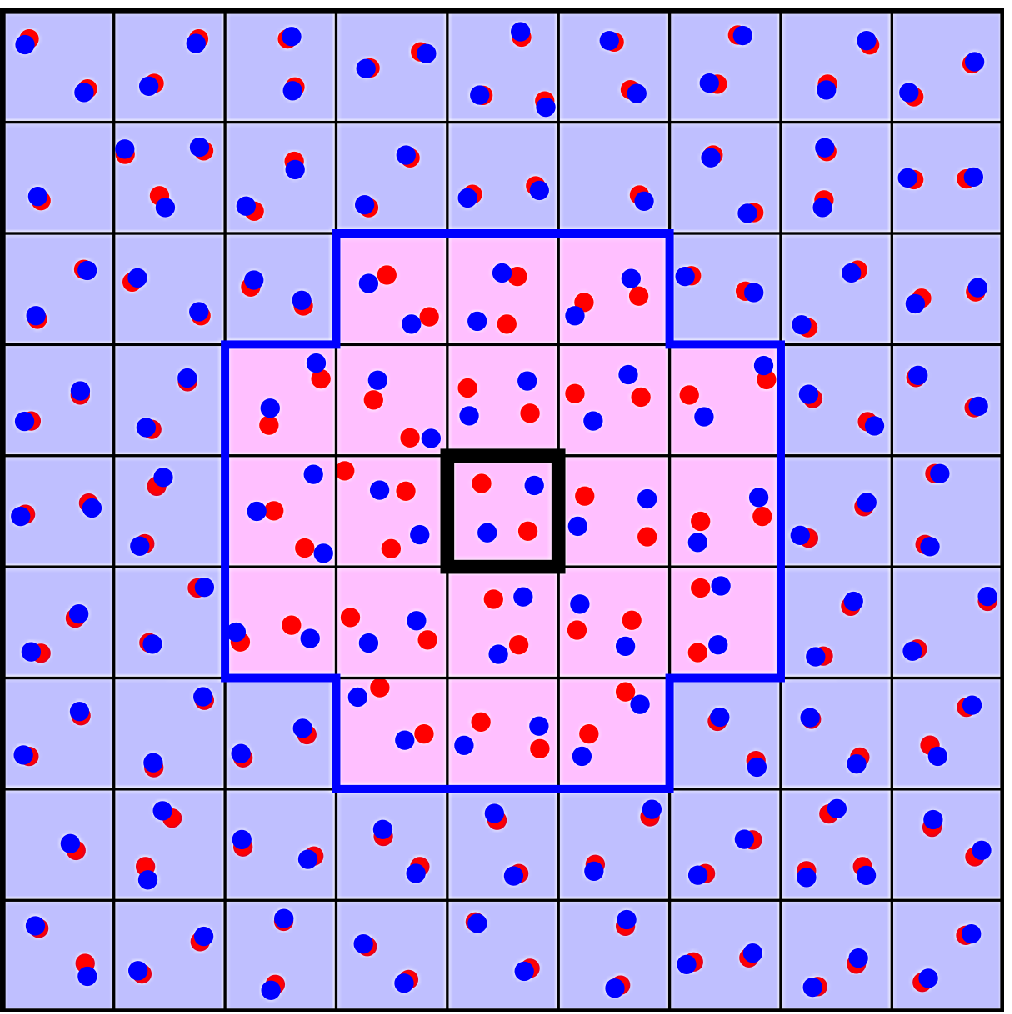}}
  \caption{The soft point-to-set thought experiment.}
  \label{BBY}
\end{figure*}

We now define a variant of the point-to-set correlation function given above, which retains desired qualitative features while admitting smooth passage to an effective-theoretic description.
Specifically, let us divide the entire system into a regular lattice of hypercubic cells (Fig.\ref{BBY1}) and, as before, fix all the particles outside some region ${\cal R}$ (Fig.\ref{BBY2}).
Then, rather than requiring a new configuration ${\bf X}_2$ to exactly match the original configuration ${\bf X}_1$ outside ${\cal R}$, we demand it to overlap highly with the original (Fig.\ref{BBY3}).
We can implement this constraint by replacing the ``hard" conditional function given in Eq.(\ref{hard}) by a ``soft" conditional function
\be\label{soft}
{\cal C}^{\rm soft}_{\overline{\cal R}}\le({\bf X}_2\Big|{\bf X}_1\ri)\equiv\prod_{\overline{\cell}\subset\overline{\cal R}}\theta\le(\hat{q}\le(\overline{\cell}; {\bf X}_1, {\bf X}_2\ri)-\qth\ri),
\ee
where the step function $\theta(x)=1$ for $x>0$ and zero otherwise~\cite{constraint}.
For sufficiently supercooled liquids, this soft constraint permits vibrations of particles around their itinerant centers; meanwhile, we choose the threshold value $\qth$ high enough so as to penalize, outside the region ${\cal R}$, significant cooperative rearrangements~\cite{AG} of these centers.
Henceforth, we work with a soft point-to-set correlation function
\bea\label{sPTS}
&&G_{\rm soft}(\cen; \overline{\cal R})\equiv\frac{1}{Z}\int d{\bf X}_1 e^{-\beta H({\bf X}_1)}\\
&\times&\le\{\frac{\int d{\bf X}_2 e^{-\beta H({\bf X}_2)}{\cal C}^{\rm soft}_{\overline{\cal R}}\le({\bf X}_2\Big|{\bf X}_1\ri)\hat{q}\le(\cen; {\bf X}_1, {\bf X}_2\ri)}{\int d{\bf X} e^{-\beta H({\bf X})}{\cal C}^{\rm soft}_{\overline{\cal R}}\le({\bf X}\Big|{\bf X}_1\ri)}\ri\}.\nonumber
\eea

As it stands, the integrand of the soft point-to-set correlation function holds $\int d{\bf X} e^{-\beta H({\bf X})}{\cal C}^{\rm soft}_{\overline{\cal R}}\le({\bf X}\Big|{\bf X}_1\ri)\equiv Z^{\rm soft}_{\overline{\cal R}}\le({\bf X}_1\ri)$ as its denominator.
This factor obstructs us from integrating out the microscopic variable ${\bf X}_1$, which is an essential formal step in passing to an effective-theoretic description.
The replica trick surmounts this obstacle.
Namely, we
introduce $\le(\nr-2\ri)$ replicas ${\bf X}_3,...,{\bf X}_{\nr}$ and insert
\be
1=\frac{\int \le\{\prod_{b=3}^{\nr}d{{\bf X}}_{b}e^{-\beta H({\bf X}_{b})}{\cal C}^{\rm soft}_{\overline{\cal R}}\le({\bf X}_b\Big|{\bf X}_1\ri)\ri\}}{\le\{Z^{\rm soft}_{\overline{\cal R}}\le({\bf X}_1\ri)\ri\}^{(\nr-2)}}
\ee
into the integrand.
In particular the denominator of the integrand now becomes $Z^{\rm soft}_{\overline{\cal R}}\le({\bf X}_1\ri)\times\le\{Z^{\rm soft}_{\overline{\cal R}}\le({\bf X}_1\ri)\ri\}^{\nr-2}=\le\{Z^{\rm soft}_{\overline{\cal R}}\le({\bf X}_1\ri)\ri\}^{\nr-1}$ and thus can be eliminated by taking the
replica limit $\nr\rightarrow1$.
Also noting $Z=\lim_{\nr\rightarrow1}Z^{\nr}$, we obtain
\bea\label{replicated}
G_{\rm soft}(\cen; \overline{\cal R})
&=&\lim_{\nr\rightarrow1}\Bigg[\frac{1}{Z^{\nr}}\int \le\{\prod_{a=1}^{\nr}d{{\bf X}}_{a}e^{-\beta H({\bf X}_{a})}\ri\}\\
&\times&\le\{\prod_{b=2}^{\nr}{\cal C}^{\rm soft}_{\overline{\cal R}}\le({\bf X}_b\Big|{\bf X}_1\ri)\ri\}\hat{q}\le(\cen; {\bf X}_1, {\bf X}_2\ri)\Bigg].\nonumber
\eea

From here, we can just turn the crank of the standard coarse-graining procedure.
For each pair of replica configurations ${\bf X}_a$ and ${\bf X}_b$ with $a\ne b$, we define a mutual local overlap at each cell $\cell$ by [c.f.~Eq.(\ref{overlap}) in the Appendix]
\be
\hat{q}_{ab}\le(\cell\ri)\equiv \hat{q}\le(\cell; {\bf X}_a, {\bf X}_b\ri)=\hat{q}_{ba}\le(\cell\ri).
\ee
We then insert
\be
1=\int \le[{\cal D}q\ri] \le\{\prod_{a<b} \prod_{\cell}  \delta\le(q_{ab}\le(\cell\ri)-\hat{q}_{ab}\le(\cell\ri)\ri)\ri\},
\ee
with the functional integral
\be
\int \le[{\cal D}q\ri][...]\equiv\int \prod_{a<b} \prod_{\cell} \le\{dq_{ab}\le(\cell\ri)\ri\}[...],
\ee
into the integrand of the replica expression given in Eq.(\ref{replicated}).
We obtain
\bea
G_{\rm soft}(\cen; \overline{\cal R})
&=&\lim_{\nr\rightarrow1}\Bigg[\int \le[{\cal D}q\ri]{\rm exp}\le(-\beta{\cal H}_{\nr}\le[q\ri]\ri)\\
&\times&\le\{\prod_{b=2}^{\nr}\prod_{\overline{\cell}\subset\overline{\cal R}}\theta\le(q_{1b}\le(\overline{\cell}\ri)-\qth\ri)\ri\}q_{12}\le(\cen\ri)\Bigg]\nonumber
\eea
where the effective Hamiltonian ${\cal H}_{\nr}\le[q\ri]$ is implicitly defined through
\bea\label{EFT}
e^{-\beta{\cal H}_{\nr}\le[q\ri]}
&\equiv&\frac{1}{Z^{\nr}}\int \le\{\prod_{a'=1}^{\nr}d{{\bf X}}_{a'}e^{-\beta H({\bf X}_{a'})}\ri\}\ \ \ \\
&\times&\le\{\prod_{a<b} \prod_{\cell}  \delta\le(q_{ab}\le(\cell\ri)-\hat{q}_{ab}\le(\cell\ri)\ri)\ri\}.\nonumber
\eea
More concisely,
\bea
G_{\rm soft}(\cen; \overline{\cal R})=\lim_{\nr\rightarrow1}\int \le[{\cal D}q\ri]\Big|_{{\overline{\cal R}}{\rm-soft}}\!\!\!\!\!\!\!\!\! e^{-\beta{\cal H}_{\nr}\le[q\ri]}q_{12}\le(\cen\ri).
\eea
Here, the constraint ${\overline{\cal R}}$-soft indicates that the
field components $q_{12},...,q_{1\nr}$ are constrained to take values greater than the threshold value $\qth$ everywhere outside the region ${\cal R}$.

Having expressed the soft point-to-set correlation function in the language of the effective overlap field theory, let us perform the thought experiment for one last time.
For sufficiently supercooled liquids, we assume that the field theory possesses, in addition to the low-overlap stable state, at least one high-overlap metastable state satisfying $q_{12},...,q_{1\nr}>\qth$ throughout the space~\cite{meta}.
Now, even when the size of the region ${\cal R}$ is very small, there exists a small droplet configuration interpolating high-overlap values outside ${\cal R}$ and a low-overlap value at its core $\cen$.
However, it is subdominant to the metastable configuration with, in particular, high $q_{12}\le(\cen\ri)$.
Thus we expect a high point-to-set correlation.
When ${\cal R}$ is very large, on the contrary, large droplet configurations dominate over the metastable configuration, resulting in a low point-to-set correlation.
The crossover takes place when the size of the region ${\cal R}$ crosses the size of critical droplets, in other words, instantons.
Therefore the size of the instantons corresponds to the point-to-set correlation length of the supercooled liquid.

We elucidated a relation between point-to-set correlations in supercooled liquids and instantons in an effective overlap field theory.
It would be interesting to thoroughly explore instantons with intricate replica-symmetry breaking patterns, building on pioneering work in~\cite{F,DSW}.
The other intriguing avenue of pursuit would be to clarify the role of point-to-set correlations in the dynamics of supercooled liquids.
For example, Montanari and Semerjian proved rigorous bounds between a point-to-set correlation length and a relaxation time for graphical models: it would be valuable to adapt their proof for generic many-body systems.
More ambitiously, it would be exciting to
tie the point-to-set correlation length to the size of dynamically heterogeneous patches~\cite{Ediger,DHbook}.

The author would like to thank Ludovic~Berthier, Ethan~S.~Dyer, and Jaehoon~Lee for discussions.

\appendix
\section{APPENDIX}
We can quantify a local overlap between two configurations ${\bf X}=\le\{{\bf x}_i\ri\}_{i=1,...,N}$ and ${\bf Y}=\le\{{\bf y}_i\ri\}_{i=1,...,N}$ within a hypercubic cell $\cell$ by
\be\label{overlap}
\hat{q}\le(\cell; {\bf X}, {\bf Y}\ri)\equiv\frac{1}{\rho \block^d}\Bigg[\frac{1}{2}\le(\sum_{\begin{subarray}{c}i,j=1\\ {\bf x}_i\in\cell\end{subarray}}^{N}+\sum_{\begin{subarray}{c}i,j=1\\ {\bf y}_j\in\cell\end{subarray}}^{N}\ri)w\le(|{\bf x}_i-{\bf y}_j|\ri)\Bigg].
\ee
Here, $\rho=\frac{N}{V}$ is a density, $l$ is the linear size of the cell $\cell$, and $w(z)$ is a monotonically decreasing short-ranged function with $w(0)=1$; we choose the size $l$ and the range of the function $w(z)$ to be of the order of the average interatomic distance.
Given these choices, for identical configurations ${\bf X}={\bf Y}$, the local overlap $\hat{q}\le(\cell; {\bf X}, {\bf X}\ri)$ takes a value close to $1$ on average.
On the other hand, for two statistically decorrelated configurations, the local overlap on average takes a low value, given in~\cite{asymptotic}.


\begin{thebibliography}{11}
\bibitem{BBreview}
For a review, see
  L.~Berthier and G.~Biroli,
  Rev.\ Mod.\ Phys.\  {\bf 83}, 587 (2011).
\bibitem{BB} 
  J.-P.~Bouchaud and G.~Biroli,
  J.\ Chem.\ Phys.\ {\bf 121}, 7347 (2004).
\bibitem{KTW}
  T.~R.~Kirkpatrick, D.~Thirumalai, and P.~G.~Wolynes,
  Phys.\ Rev.\ A\ {\bf 40}, 1045 (1989).
\bibitem{MSbound}
 A.~Montanari and G.~Semerjian,
  J.\ Stat.\ Phys.\ {\bf 125}, 22 (2006).
\bibitem{DSW}
  M.~Dzero, J.~Schmalian, and P.~G.~Wolynes
  Phys.\ Rev.\ B\ {\bf 72}, 100201 (2005).
\bibitem{CBTT}
  C.~Cammarota, G.~Biroli, M.~Tarzia, and G.~Tarjus,
  Phys.\ Rev.\ Lett.\  {\bf 106}, 115705 (2011).
\bibitem{F}
  S.~Franz,
  J.\ Stat.\ Mech.\ P04001 (2005).
\bibitem{FM}
  S.~Franz and A.~Montanari,
  J.\ Phys.\ A\ {\bf 40}, F251 (2007).
\bibitem{BBCGV}
 G.~Biroli, J.-P.~Bouchaud, A.~Cavagna, T.~S.~Grigera, and P.~Verrocchio,
 Nature\ Phys.\  {\bf 4}, 771 (2008).
\bibitem{HMR}
  G.~M.~Hocky, T.~E.~Markland, and D.~R.~Reichman,
  Phys.\ Rev.\ Lett.\ {\bf 108}, 225506 (2012).
\bibitem{AG}
  G.~Adam and J.~H.~Gibbs,
  J.\ Chem.\ Phys.\ {\bf 43}, 139 (1965).
\bibitem{B}
  L.~Berthier,
  Phys.\ Rev.\ E\ {\bf 88}, 022313 (2013).
\bibitem{Ediger}
  M.~D.~Ediger,
  Annu.\ Rev.\ Phys.\ Chem.\ {\bf 51}, 99 (2000).
\bibitem{DHbook}
  L.~Berthier, G.~Biroli, J.-P.~Bouchaud, L.~Cipelletti, and W.~van~Saarloos, (eds.),
  {\it Dynamical Heterogeneities in Glasses, Colloids, and Granular Media}
  (Oxford University Press, Oxford, 2011).
\bibitem{Franzology}
See~\cite{F,FM} for an analogous line of work on a spherical Kac $p$-spin glass model.
\bibitem{details}
For simplicity, we work with the canonical ensemble and classical (as opposed to quantum) particles.
For notational simplicity, we ignore momenta ${\bf P}=\le\{{\bf p}_i\ri\}_{i=1,...,N}$ and an irrelevant normalization factor: if wished, replace $\int d{\bf X}[...]$ by $\frac{1}{N! (2\pi\hbar)^{dN}}\int d{\bf X} d{\bf P}[...]$.
When we partition the system into a region ${\cal R}$ and its complement $\overline{\cal R}$, we must sum over $N_{\cal R}$, the number of particles inside ${\cal R}$;
alternatively, we can work with the grandcanonical ensemble throughout.
\bibitem{asymptotic}
The asymptotic low-overlap value of the point-to-set correlation functions in the large region limit is given by
\be
\int d{\bf X}_1 {\cal M}({\bf X}_1)\int d{\bf X}_2 {\cal M}({\bf X}_2)\hat{q}\le(\cen; {\bf X}_1, {\bf X}_2\ri),\nonumber
\ee
where, without loss of generality, we can place a cell $\cen$ anywhere in the system.
We may define ``connected" point-to-set correlation functions by subtracting this value from the ``disconnected" point-to-set correlation functions given in Eq.(\ref{PTS}) and in Eq.(\ref{sPTS}).
\bibitem{constraint}
The specific form of this soft conditional function is immaterial.
Crucial points are that it is a function of local overlaps and that it penalizes low overlaps outside the region ${\cal R}$.
\bibitem{meta}
This is customarily assumed (for example, see~\cite{DSW,CBTT}).
See~\cite{B} for interesting numerical work which can potentially access these metastable states directly.
\end{thebibliography}
\end{document}